\documentclass{article}

\usepackage{microtype}
\usepackage{graphicx}
\usepackage{subfigure}
\usepackage{booktabs} 

\usepackage{hyperref}

\usepackage[accepted]{socialgood}

\icmltitlerunning{Open Platforms for Artificial Intelligence for Social Good}

\begin{document}

\twocolumn[
\icmltitle{Open Platforms for Artificial Intelligence for Social Good: \\Common Patterns as a Pathway to True Impact}



\icmlsetsymbol{equal}{*}

\begin{icmlauthorlist}
\icmlauthor{Kush R.\ Varshney}{ibm}
\icmlauthor{Aleksandra Mojsilovi\'{c}}{ibm}
\end{icmlauthorlist}

\icmlaffiliation{ibm}{IBM Research, Yorktown Heights, NY, USA}

\icmlcorrespondingauthor{Kush R.\ Varshney}{krvarshn@us.ibm.com}

\icmlkeywords{Machine Learning, ICML}

\vskip 0.3in
]



\printAffiliationsAndNotice{}  

\begin{abstract}
The AI for social good movement has now reached a state in which a large number of one-off demonstrations have illustrated that partnerships of AI practitioners and social change organizations are possible and can address problems faced in sustainable development. In this paper, we discuss how moving from demonstrations to true impact on humanity will require a different course of action, namely open platforms containing foundational AI capabilities to support common needs of multiple organizations working in similar topical areas.  We lend credence to this proposal by describing three example patterns of social good problems and their AI-based solutions: natural language processing for making sense of international development reports, causal inference for providing guidance to vulnerable individuals, and discrimination-aware classification for supporting unbiased allocation decisions.  We argue that the development of such platforms will be possible through convenings of social change organizations, AI companies, and grantmaking foundations.
\end{abstract}

\section{Introduction}
\label{sec:intro}

Since its incipient stages seven or eight years ago \cite{Porway2011,CatlettG2015,Varshney2017}, the data science and artificial intelligence (AI) for social good movement has created a very large portfolio of examples that show how partnerships between (a) technologists, and (b) non-profits, social enterprises, government agencies or other similar mission-driven organizations can touch on the seventeen sustainable development goals (SDGs) \cite{ChuiHMRCHN2018}, which include poverty, hunger, ill health, and inequalities of various sorts \cite{NilssonGV2016}.  These are the most difficult of problems because they involve so many different issues and so many different stakeholders. 

Striving to achieve the global goals includes not only technical challenges, but a large number of \emph{socio}-technical challenges that require the collaboration of many different parties with varying expertise and imagination of what is possible.  Thus far, the AI for good movement has increased the AI literacy of social change organizations and shown how to formulate pain points in sustainable development as tangible and tractable AI problems. The breaking down of collaboration barriers is already remarkable, but should only be seen as a foundation or `Act I' of a story yet to come.

The movement has included several models of engagement: data science competitions (e.g.\ DrivenData and Data Science Bowl), weekend volunteer events (e.g.\ DataKind DataDives and hackathons), longer term volunteer-based consulting projects (e.g.\ DataKind DataCorps), fellowship programs (e.g.\ Data Science for Social Good and IBM Science for Social Good), corporate philanthropy (e.g.\ Two Sigma Data Clinic), specialized non-governmental organizations (e.g.\ Bayes Impact), innovation units within large development organizations (e.g.\ United Nations Global Pulse), and data scientists employed directly by smaller social change organizations.  Irrespective of the engagement model, nearly all efforts and projects have been small custom-tailored demonstrations or case studies \cite{ChuiHMRCHN2018,HowsonCIB2018}. Arguably, there has not been a truly measurable and lasting impact from any project in the broader portfolio.

The most significant bottlenecks identified by \citet{ChuiHMRCHN2018} that have led to this state of affairs are: data inaccessibility, shortages of talent, and `last mile' implementation challenges. In the remainder of this paper, we focus on the latter two of these three limiting factors and ask whether a continuation of the custom-tailored or bespoke approach taken thus far is what can take us to `Act II,' or whether another approach may be required.  We discuss why a \emph{platform}-based approach that takes advantage of common patterns in AI formulations for the social good space may be the preferred route --- as long as it is pursued in an open way and with continued partnerships between the technology and social sectors that have been essential to the successes of Act I.

After discussing the broad idea of open platforms for AI for social good, we will delve into three specific patterns of problems we have seen: making sense of international development and humanitarian crisis reports, providing guidance for vulnerable individuals, and supporting unbiased allocation decisions.  Due to page limits, our aim is not to be comprehensive, but to lend credence to the proposal by illustrating that there are in fact themes containing enough common required functionality to be candidates for an AI platform.  We could have, of course, discussed other patterns and other projects within those patterns herein. We have chosen ones that are closer to the cutting edge of machine learning research and ones we have personally worked on as founding co-directors of the IBM Science for Social Good initiative, and as data ambassadors with DataKind.

\section{Proposal of Open Platforms for AI for Social Good}
\label{sec:proposal}

As we discussed in Section \ref{sec:intro}, the current modus operandi of the AI for good movement to do customized one-off projects has been successful in starting the dialogue between the social sector and the world of AI research and development, which is in and of itself an important accomplishment.  However, let us pause and consider whether this mode is in humanity's best interest or whether there may be alternatives to go forward with.

First let us recount a few realities of the current way of doing things.  One-off solutions require a great deal of time and effort in creating, both from members of the social change organization and from the data scientists involved --- which neither have in abundance. There is very limited ability to reuse assets and learnings from one project to the next since every new project involves a different organization, and volunteer data scientists or fellows are often unable to commit to conducting several projects over time.  Also, custom solutions require the social change organization to integrate, deploy, monitor, and maintain the work product, which even the best organizations struggle to do because of skill and resource limitations. 

\citet{Burgess2018} has identified three main categories for creating AI solutions: off-the-shelf AI software, AI platforms, and bespoke AI builds.  He states that ``the biggest disadvantage of using [off-the-shelf AI software] is that its capabilities may not align well enough with your objectives and required functionality.''  This is clearly true for social good problems because they do tend to have some unique characteristics as compared to problems in other industries and sectors.  He continues that ``for the majority of enterprises, bespoke AI development should be used only when absolutely necessary, that is, for complex, very large data problems, or when creating a completely new product or service that requires technological competitive advantage.''  Although bespoke AI builds are the current practice in AI for social good, the typically-encountered problems actually have small datasets and do not require a technological competitive advantage; the core innovation in AI for social good is typically not on the very fine technical details, but on the application, adaptation, and combination of technologies to the real-world problem.  

If not off-the-shelf and if not bespoke, we are left with AI platforms.  AI platforms provide sets of foundational capabilities that can be further specialized for particular instantiations or problems, trained on users' datasets, and deployed and maintained with relative ease via cloud computing environments.  Reusability of the foundational AI capability is inherent to platforms because the algorithms are created once and can be improved over time by a dedicated team in a way that diffuses to deployments automatically.  A given AI platform is used by many different organizations who tend to have similar, but not exactly the same needs.  This mode makes sense for AI for social good because it significantly reduces the expertise and resource requirements of the social change organization while also allowing AI developers, who themselves are scarce, to multiply their impact by working on the core functionality that may then be used by many mission-driven organizations.

AI platforms can be highly generic and general or they can be more specific to particular industries or functions.  For example, AI companies have started creating platform offerings specific to industries such as agriculture, manufacturing and building management, as well as functions such as customer service, human resources and advertising \cite{IBMa}.  To the best of our knowledge, however, there is no such AI platform for the sorts of problems typically encountered in sustainable development and social good.  We believe that this gap reveals a call to action for building platforms for clusters of problems as a central activity of Act II of the AI for social good movement. 

How will this happen? There must be a convening of three categories of parties: social change organizations, (private) AI companies, and grantmaking foundations.  To create a platform for a cluster or common pattern of problems, it is essential that a good number of social change organizations participate because a platform created without the input and feedback of the eventual users will necessarily be suboptimal. The collection of organizations should be somewhat diverse but also have similar AI needs in order to stretch the boundaries of the design and capabilities while ensuring that they are general enough for all use cases. (This is actually just a best practice of user-centered design \cite{Guszcza2018}.)

AI companies are needed in the convening because they hold the required skills and talents to build such platforms.  They can also leverage internal tools and practices to make the endeavor less daunting.  Individual developers working on a volunteer basis are unlikely to be able to support such an undertaking, at least in its initial development.  By creating AI platforms for social good, the technology solution creators accrue an additional benefit by learning lessons applicable for developing platforms in other domains.

\begin{figure*}[ht]
\vskip 0.2in
\begin{center}
\centerline{
\begin{tabular}{ccc}
\includegraphics[height=0.3\textwidth]{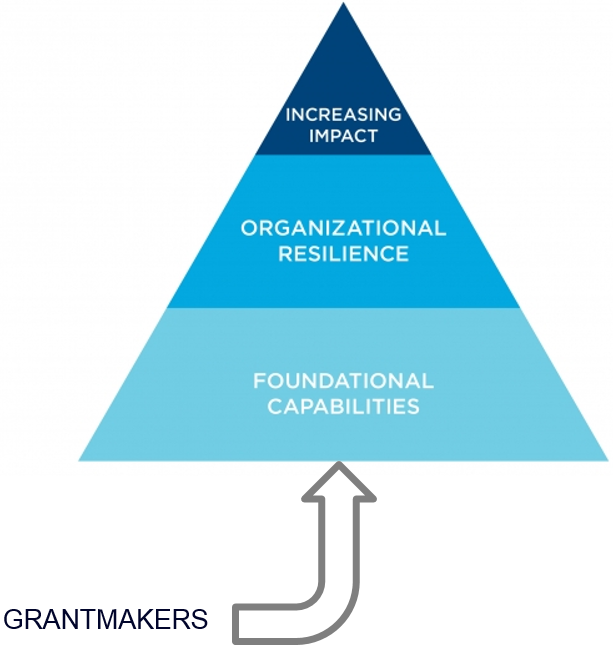} & \includegraphics[height=0.3\textwidth]{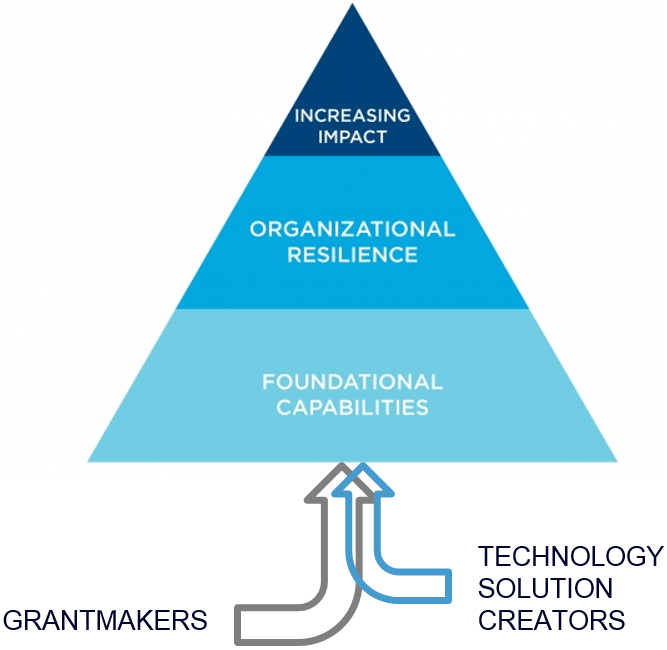} & \includegraphics[height=0.3\textwidth]{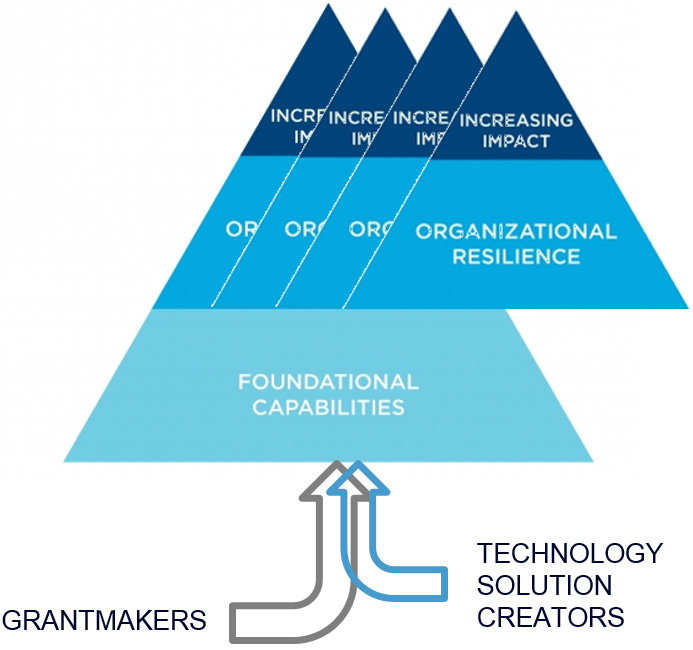} \\
\footnotesize{(a)} & \footnotesize{(b)} & \footnotesize{(c)}
\end{tabular}
}
\caption{(a) Grantmaking foundations are starting to understand that they cannot support only programmatic efforts of social change organizations; they must also support foundational capabilities to enable sustained impact. (b) Even if grantmakers want to support new solutions and foundational technologies, social change organizations are not in position to create these solutions themselves. Technology solution creators must also be involved. (c) The same platform can serve as a foundational capability to support not just one, but many social change organizations with similar missions. (This figure is a modification of a figure by \citet{EtzelP2017}).}
\label{fig:pyramids}
\end{center}
\vskip -0.2in
\end{figure*}

Finally, grantmaking foundations are needed to pull together social change organizations working in similar areas (many foundations already support organizations grouped by topical area), give them `permission' to participate in such an activity that does not immediately advance their mission, and to provide market validation to participating AI companies.  Grantmakers are already starting to understand that they cannot continue their typical existing practice of earmarking funds solely for programmatic efforts; they must also support foundational capabilities to enable sustained impact \cite{EtzelP2017}.  The overall description of the convening of grantmakers, AI companies, and social change organizations is illustrated in Figure \ref{fig:pyramids}. 

The general idea has seemingly started to take root.  In January 2019, the Rockefeller Foundation and the Mastercard Center for Inclusive Growth announced an investment that ``will enable DataKind to move from a project- to a platform-based model, in which it will support more organizations on a set of common issues'' \cite{Shah2019}.

Before closing this section and describing a few examples of sets of social change organizations with similar AI needs, let us bring forth a discussion of \emph{openness} in the platforms.  As we have described platforms thus far, we have implied that they enable \emph{linear} scaling.  That is, due to reuse of the core AI capability in the platform, more and more social change organizations will have their programmatic efforts supported with the same amount of AI development effort.  However, when platforms are open, they also enable new innovations, which has time and again yielded \emph{exponential} scaling of impact \cite{GawerC2013,AhujaC2016,BogersCM2018}.  Open-source toolkits are already a key component of the fabric of AI research and development, so it is not unreasonable to desire that platforms for AI for social good be open from the outset --- to minimize costs to social change organizations who use them, to allow volunteers to add capabilities, \emph{and} to unlock greater innovation.

\section{Making Sense of International Development and Humanitarian Reports}
\label{sec:development}

The International Accountability Project (IAP) is an international advocacy organization that, through its Early Warning System, collects information about planned and early-stage development projects such as geothermal power plants, irrigation projects and the privatization of early childhood education supported by institutions such as the World Bank, the Asian Development Bank and the European Investment Bank.  It uses this information to alert the civil society of the local affected population of the plans and supports them in flagging potential problems and proposing alternatives that are more in line with the priorities of the local population.

One of us (Varshney) led a group of volunteer data scientists at a DataKind DataDive aiming to improve and automate the Early Warning System \cite{Bert2018}, which at that point was primarily a manual undertaking performed by over-burdened analysts. The main tasks were to take articles from news feeds, categorize them according to topic, extract metadata such as the time, place and development bank, and match them to known development projects using semantic natural language processing (NLP)-based analysis.  By automating much of this work, analysts would be freed to perform higher-level reasoning about the projects.  

We have seen the same pattern in several projects we have conducted in the IBM Science for Social Good initiative.  Analysts at ACAPS collect and summarize secondary information about humanitarian crises (news reports and reports from various humanitarian agencies) to help prioritize responses; the organization has more-or-less the same AI and NLP needs as IAP \cite{PhamSDJVCFMV2017}.  The Clinton Global Initiative, when it was still in existence, wished to analyze the corpus of commitment reports made by its members to conduct large philanthropic and development projects. The purpose was to provide guidance for future projects; this work also required similar topic analysis of natural language text and extraction of metadata as for IAP, coupled with more advanced recommender system algorithms \cite{LambaHSLBBMV2017}.  Similar to IAP, the United Nations Development Programme (UNDP) and the International Center for Advocates Against Discrimination (ICAAD) sought out NLP algorithms to perform semantic matching of given snippets of text to target text, in these cases matching sentences of national development plans (UNDP) and items from the United Nations' Universal Periodic Review of Human Rights (ICAAD) to individual targets of the SDGs \cite{ICAAD2016,ICAAD2017,GalsurkarSWVSIKV2018}.

\section{Providing Guidance for Vulnerable Individuals}
\label{sec:vulnerable}

CityLink Center is a provider of integrated social services in Cincinatti, USA. Through coordinated interventions such as one-on-one counseling sessions on different topics and group educational classes, it aims to help its clients get out of poverty.  It records progress along several realms of life such as employment, housing, transportation, education, and mental health.  

In an ongoing project we are conducting through the IBM Science for Social Good initiative, CityLink's longitudinal event data of interventions and outcomes is being analyzed to discover the causal links among the various educational sessions and realms of life. This model can then be used to provide personalized guidance on future interventions to maximize the probability of a client graduating out poverty or to minimize the time taken. 

We partnered with Neighborhood Trust Financial Partners, a financial empowerment services provider to low-wage workers in New York, USA, to improve their WageGoal app that uses data on spending and income transaction events to provide personalized predictions on when clients may go into debt and provides guidance on actions to prevent this from happening \cite{ZhangHRWYAV2018}.  We also use personalized causal inference in our work on analyzing medical claims data to provide guidance on opioid prescribing to prevent adverse outcomes such as addiction and long-term use which may lead to overdose deaths \cite{ZhangIWVBMFYMV2017,WeiJ2018}. St.\ John's Bread and Life, a provider of emergency food and other services in Brooklyn, USA and the International Center for Appropriate and Sustainable Technology, a non-profit retrofitter of multifamily affordable housing for energy conservation in Colorado and other states of the USA have similar AI needs as CityLink of understanding client behavior through causal event modeling and intervention optimization \cite{HelanderMKMB2018,IBMb}.  

\section{Supporting Unbiased Allocation Decisions}
\label{sec:fairness}

Small-amount mobile money-based loans in East Africa, such as M-Shwari, are now approved or declined in a pipeline involving machine learning classifiers \cite{SpeakmanSM2018}.  In a recent project of ours, \citet{CostonRWVSMC2019} examined the fairness of these algorithms with respect to potentially protected attributes such as gender and age, and developed algorithms to mitigate unwanted biases in these allocation decisions that have a profound effect on the typically poor and traditionally-unbanked applicants.  

Similarly, our DataKind DataCorps project in partnership with Simpa Networks, a pay-as-you-go solar panel system provider in rural India, developed a machine learning classifier as an approval mechanism to predict which applicants will not fully repay for their systems and will have to have their systems repossessed \cite{GerardRSVKN2015}. A potential feature for the task, the surname of the applicant, which strongly correlates with caste and religion, was explicitly excluded to prevent unwanted bias \cite{VarshneyA2017}.  GiveDirectly, a provider of unconditional cash transfers, participated in a DataCorps project to prioritize villages in western Kenya to be allocated donations based on poverty levels estimated from satellite imagery \cite{AbelsonVS2014}. In our post hoc analysis of the first pilot, we found unwanted biases resulting from inappropriate consideration of buildings in housing compounds that were not the main house \cite{VarshneyCANSXS2015}. A slightly different example that fits the AI pattern was our IBM Science for Social Good project conducted with Echoing Green, a funder of social entrepreneurs.  Classifiers to predict fellowship candidates' advancement to the semifinalist stage was pursued and potential biases by gender, country of origin, and educational institution were examined \cite{GargOSKKMMBV2017}.

\section{Conclusion}
\label{sec:conclusion}

The most challenging problems facing humanity today are succinctly captured by the SDGs, and all of them require the combination of multiple kinds of expertise to address.  The data science and AI for social good movement has been working for the last several years to partner AI practitioners with social change organizations to show the potency of this combination.  Dialogue among these groups is no longer completely foreign due to the efforts thus far, but to make measurable progress towards the goals, the current mode of conducting bespoke projects needs to be enhanced by the creation of open AI platforms.  

Platforms will provide reusable base AI capabilities for common sets of problems faced by groups of social change organizations in a way that is customizable, deployable, and maintainable with as little effort as possible.  In doing so, they will overcome the bottlenecks of AI talent shortages and last mile implementation while also enabling further innovation beyond the missions of the participating organizations.  We feel that the way to achieve such progress is through consortia of three parties: social change organizations, AI companies, and grantmaking foundations, and can already imagine several initial candidate social good platforms through the examples discussed herein.

The third bottleneck identified by \citet{ChuiHMRCHN2018} is data accessibility.  We recognize that this is an additional challenge, but also posit that there is enough relevant open data and data owned by social change organizations to address their missions already if open innovation platforms were available to them \cite{KapoorMSV2015}.  Corporate data philanthropy should catch up soon enough.  Moreover, we recognize that scaling is not the mission of many social change organizations \cite{Slaughter2018}; however, it is not the scaling of the organizations themselves that we are advocating in this work.  We are advocating for the creation of scalable core AI capabilities that will make the work of all organizations easier, whether they would like to remain small with a very pointed mission or to grow.

\section*{Acknowledgements}
The authors thank Modest Oprysko for discussions and Jake Porway for introducing them to the AI for social good movement.  They also thank Dario Gil, John Kelly, Arvind Krishna, Mahmoud Naghshineh, Giovanni Pacifici, Radha Ratnaparkhi, and Bob Sutor for supporting the IBM Science for Social Good initiative and its 36 fellows to-date, the IBM scientists and engineers who have participated in projects, and social change organization partners.

\bibliography{socialgood}
\bibliographystyle{icml2019}

\end{document}